\documentclass[amsmath,amssymb,reprint]{revtex4-1}
\DeclareMathAlphabet\mathbfcal{OMS}{cmsy}{b}{n}

\usepackage{blindtext}
\usepackage{amsmath}
\usepackage{siunitx}
\usepackage{graphicx}
\usepackage{bm}

\usepackage{hyperref}
\hypersetup{
	colorlinks=true,
	linkcolor=blue,    
	urlcolor=blue,
}

\usepackage[capitalise]{cleveref}

\begin{document}
\title{Efficient ensemble uncertainty estimation in Gaussian Processes Regression}

\author{Mads-Peter Verner Christiansen}
\author{Nikolaj Rønne}
\author{Bjørk Hammer}
    \email{hammer@phys.au.dk}
\affiliation{Center for Interstellar Catalysis, Department of Physics and Astronomy, Aarhus University, DK‐8000 Aarhus C, Denmark}

\date{\today}

\begin{abstract}
Reliable uncertainty measures are required when using data based
machine learning interatomic potentials (MLIPs) for atomistic simulations. In
this work, we propose for sparse Gaussian Process Regression type MLIP a
stochastic uncertainty measure akin to the 
query-by-committee approach often used in conjunction with
neural network based MLIPs. The uncertainty measure is coined
\textit{``label noise"} ensemble uncertainty as it emerges from adding noise
to the energy labels in the training data. We
find that this method of calculating an ensemble uncertainty is as well calibrated as the one
obtained from the closed-form expression for the posterior variance
when the sparse GPR is treated as a projected process. Comparing the
two methods, our proposed ensemble uncertainty is, however, 
faster to evaluate than the closed-form expression. Finally, we demonstrate that the proposed 
uncertainty measure acts better to support a Bayesian search for optimal structure of
Au$_{20}$ clusters.
\end{abstract}

\maketitle

Machine learning interatomic potentials (MLIPs) based on density functional theory
(DFT) data are currently replacing full-blown DFT studies in computational
materials science and chemistry. The seminal works by Behler and
Parrinello \cite{behler2007} and Bartok et.\ al \cite{bartok2010}, led the way to this transition by introducing how MLIPs can be
implemented using either a neural network (NN) or a
Gaussian Process Regression (GPR) approach. Since those works, the
field has seen significant improvements and nowadays
simulations are rutinely made for large atomistic systems. Recent examples
are MLIP-based molecular dynamics (MD) simulations of amorphous carbon and
silicon\cite{Deringer2017,Deringer2021}, for dissociative adsorption of
hydrogen over Pt surfaces \cite{Vandermause2022}, and for vibrational
spectroscopy of molecules and
nano-clusters\cite{gasteggerMachineLearningMolecular2017,
  Tang2023}. MLIPs are also being used in structural
search, where they significantly enhance the likelihood of identifying
e.g.\ the shape of metal clusters supported on oxide
surfaces \cite{Kolsbjerg2018, Behler2020} or where they speed up the search
for crystal surface reconstructions \cite{timmermannMathrmIrOSurface2020, wangGeneralizableMachineLearning2021, timmermannDataefficientIterativeTraining2021, merteStructureUltrathinOxide2022, leeStagedTrainingMachineLearning2023} along with 
various other applications of such techniques \cite{ouyangGlobalMinimizationGold2015, tongAcceleratingCALYPSOStructure2018, arrigoniEvolutionaryComputingMachine2021, wangMAGUSMachineLearning2023}.

MLIPs can be trained to a high level of accuracy on databases of
structure-energy data points whenever given the right representation
and model architecture. Early work on the topic focused on adapting machine learning 
regression methods for the task of fitting potential energy surfaces, such as designing effective descriptors 
that encode the invariances of the Hamiltonian \cite{valleCrystalFingerprintSpace2010, behlerAtomcenteredSymmetryFunctions2011, ruppFastAccurateModeling2012,bartokRepresentingChemicalEnvironments2013, faberAlchemicalStructuralDistribution2018,drautzAtomicClusterExpansion2019, huoUnifiedRepresentationMolecules2022}
Other developments include message-passing or graph neural networks \cite{gilmerNeuralMessagePassing2017, schuttSchNetContinuousfilterConvolutional2017, xieCrystalGraphConvolutional2018, gasteigerDirectionalMessagePassing2019} 
and their equivariant counterparts \cite{thomasTensorFieldNetworks2018, andersonCormorantCovariantMolecular2019, schuttEquivariantMessagePassing2021, batznerEquivariantGraphNeural2022, batatiaMACEHigherOrder2023}. 
Another area of interest has been the inclusion of long-range effects \cite{unkePhysNetNeuralNetwork2019, behlerMachineLearningPotentials2021, anstineMachineLearningInteratomic2023}.
Clearly, there is substantial interest in both improvements to and applications of MLIPs.
With the success of MLIPs which offer gains in computational efficiency by orders of magnitude and 
the application of these to increasingly complex atomistic systems, the ability to 
assess the quality of predictions is becoming increasingly important. 
As such, one recurring concern with MLIPs is their reliability when applied to
structures and configurations that overlap little with the training data. 
While loss metrics on a test set are useful for comparing model performance, an 
accurate uncertainty measure allows for assessing the confidence of a model's 
predictions. As such, approaches for calculating uncertainties and methods of 
evaluating whether such uncertainty estimates are consistent are active topics of 
research \cite{botuMachineLearningForce2017, a.petersonAddressingUncertaintyAtomistic2017, tran_methods_2020, wenUncertaintyQuantificationMolecular2020, hu_robust_2022, rasmussenUncertainUncertaintiesComparison2023, carreteDeepEnsemblesVs2023,jorgensenCoherentEnergyForce2023, zhuFastUncertaintyEstimates2023, tanSinglemodelUncertaintyQuantification2023}.
Additionally, uncertainties can be used to guide data acquisition or to 
enhance common computational tasks such as structure optimization and exploration.
One strategy is to use an uncertainty measure to determine whether to continue 
a MLIP based simulation or to first collect new DFT data points for refining the model \cite{behlerConstructingHighdimensionalNeural2015, a.petersonAddressingUncertaintyAtomistic2017, zaverkinExploringChemicalConformational2022, vanderoordHyperactiveLearningDatadriven2023, kulichenkoUncertaintydrivenDynamicsActive2023, zaverkinUncertaintybiasedMolecularDynamics2024}.
Likewise various active learning protocols have been formulated, in which only the most promising candidate based on an MLIP
search is selected for investigation at the DFT level followed by an
retraining of the MLIP \cite{hernandez-lobatoParallelDistributedThompson2017, Todorovic2019, Bisbo2020, kaappaGlobalOptimizationAtomic2021}.

There are two frequently used uncertainty measures in atomistic
simulations. For neural network based MLIPs, the ensemble
method is frequently employed \cite{schranCommitteeNeuralNetwork2020, montes-camposDifferentiableNeuralNetworkForce2022, kahleQualityUncertaintyEstimates2022, buskGraphNeuralNetwork2023}, while for GPR models, the closed-form
expression for the posterior variance is the natural
choice. When employing the ensemble method in conjunction with neural
networks, several models of the same architecture 
are trained on the same data. The models differ since their
networks are initialized with independent random weights. When
predictions are made with the ensemble method, the mean and variance
of the expectations are then deduced from the spread of predictions from
the  models. The method is hence also referred to as \textit{query-by-committee}. 
For GPR-based MLIP models, the uncertainty can be calculated more directly,
as a closed-form expression exists for the posterior variance. This is also the
case when using \textit{sparse} GPR models, where not every local
atomic feature present in the training data is used for the
prediction. Sparse GPR models are solved as a projected process,
where a subset of the atomic features in the training data are used as
inducing points. Projected processes also have a closed-form expression
for the posterior variance, but it involves large matrices relating to
the sparsification and hence eventually become time limiting. 

When formulating ensembles of neural network models, the obvious means
to get different models is to exploit the randomness of the initial
network weights. This cannot be carried over to the GPR domain, as the
models are deterministic, but ensembles of models could be obtained
either
\begin{itemize}
\item by training individual models on separate subsets of the total training set,
\item or by varying the hyperparameters or the form of the kernel function for each model.
\end{itemize}
Such approaches have been employed \cite{deringerGaussianProcessRegression2021}. However, both suffer from the obvious drawback
that some or all aspects of their use, including sparsification, training, and
prediction, would scale linearly with the ensemble size. In the
sparsification, each model would identify different inducing points,
in the training, each model would require solving an independent set of
equations, and in the prediction, each model would require the setup
of a different kernel vector.

In this work, we present an elegant way of avoiding the most critical
parts of this linear increase in computational demand with ensemble
size. We do so by establishing each model on replicas of \textit{all data} 
with random noise added. In this way, the models can share the
sparsification step, the setting up and inversion of the kernel matrix,
and the calculation of the kernel vector for a prediction.

Generally, the formulation of GPR models involves assuming noisy labels, 
which is evidently beneficial when trained on data, such as from experiments, 
that have noisy labels. However, whenever DFT energies are the labels, there is no noise in the data,
since new DFT calculations for the same structures would reproduce the
energies exactly. Regardless, the noisy GPR formulation is typically used to ensure
numerical stability. Furthermore, limitations to the model's ability to fit the data, 
that may arise from representation deficiencies or the limit on model complexity 
imposed by the kernel function, are handled by this formalism. 
We extend this by deliberately adding noise to the training labels in a 
manner that allows defining a computational efficient ensemble of GP regressors
for uncertainty quantification. In practice, we add a normally distributed noise to the DFT
energies and train each GPR model according to the noisy data. This can
be done a number of times, and an uncertainty measure can be obtained
from the distribution of predictions by the resulting ensemble of models.


The article is outlined as follows: We first introduce the label noise
approach and argue how the cost of having several GPR models becomes
negligible when used in conjunction with sparse GPR models, where
sparsification is the major computational bottleneck, rather than solving for
each model. Next, we investigate the quality of the uncertainty
measure for the test case of Au clusters. Finally, we demonstrate the usefulness
of the label noise uncertainty measure in MLIP-enhanced structural
searches for Au$_{20}$ clusters.

\section{Results}

\subsection{Ensemble Gaussian Process Regression}

The central model we propose is an efficient ensemble formulation, as an alternative 
to the projected-process uncertainty measure that is part of the sparse GPR formalism. 
This model is based on a sparse GPR model, described in detail in the methods section \ref{sec:method_sgpr}, 
where a local energy prediction results from the expression 
\begin{equation}
	\epsilon(\mathbf{x}) = \mathbf{k}_m(\mathbf{x}) C \mathbf{E}.
	\label{eq:lgpr_c_matrix}
\end{equation}
Here the matrix $C$ involves the inversion of a matrix of local descriptor covariances. 
For neural networks an ensemble of models can be constructed simply 
by having multiple copies of the network with different randomly of initialized weights. 
However, GPR models have no randomly set initial parameters and some other way of setting up 
different models must be invoked. Choices include bootstrapping the training data, 
that is training each model with different subsets of the training set or 
selecting different hyperparameters (kernel parameters, inducing points). 
Note that both options involve calculating both a different $C$-matrix 
and a different kernel-vector $\mathbf{k}_m(\mathbf{x})$ for each model, which comes at significant 
computational cost. Instead, if the differences between the individual models
of the ensemble is limited to the $\mathbf{E}$-term, then $C$ and
$\mathbf{k}_m(\mathbf{x})$ will only 
need to be calculated once. By defining $\mathbf{N}$ as the number of atoms for each 
energy observation $\mathbf{E}$ our proposed expression for an
individual model, $k$, of the ensemble is given by first adding normally distributed 
noise to the labels of the training data:
\begin{equation}
	\mathbf{\tilde{E}}_k = \mathbf{E} +
        \mathbf{N}\odot\bm{\gamma}_k - \rho_k \mathbf{N},
\end{equation}
where $\bm{\gamma}_k \sim \mathbfcal{N}(0, \sigma_l^2)$ is random noise on each label and $\rho_k \sim \mathcal{N}(0,\sigma_p^2)$ is 
a shift for model $k$, $\odot$ denotes element-wise multiplication, and $\mathcal{N}(\mu,\sigma^2)$ represents a
normally distributed stochastic variable, with mean $\mu$ and
variance $\sigma^2$. 
With these altered labels the prediction of a local energy from model $k$ can be expressed as:
\begin{equation}
	\epsilon_k(\mathbf{x}) = \mathbf{k}_m(\mathbf{x}) C\mathbf{\tilde{E}}_k + \rho_k.
	\label{eq:gpr_ensemble}
\end{equation}
The two noise terms added are dubbed \textit{label noise}
and \textit{prior noise}, as they act directly on each label and as a
common prior, respectively. The \textit{label noise} term
specified by $\sigma_l$ is drawn independently for \textit{each
  structure} and for each model, while the \textit{prior noise} term,
specified by $\sigma_p$, is drawn for
\textit{each model} only. The two noise terms have different
functions.  The label noise makes the various models in the ensemble
associate uncertainty with \textit{known data} and thereby influence predictions 
in the neighborhood of training data, while the prior noise
makes the various models disagree on \textit{unknown data}. These noise 
terms may also be thought of as representing aleatoric and epistemic
uncertainties, respectively. 
With more data the uncertainty arising from $\sigma_p$ may be reduced as such it is 
epistemic, whereas the uncertainty produced by $\sigma_l$ will not reduce with additional 
data -- since it is aleatoric. See \cite{hullermeier_aleatoric_2021} for further discussion 
on the distinction between these two types of uncertainty. 

The mean prediction of an ensemble with $K$ models is 
\begin{equation}
	\bar{\epsilon}(\mathbf{x}) = \frac{1}{K}\sum_k^K \epsilon_k(\mathbf{x}).
	\label{eq:ensemble_mean}
\end{equation}
As $K \rightarrow \infty$ this converges to Eq. \eqref{eq:lgpr_c_matrix} so the ensemble 
retains the mean prediction of a GP model without the added noise terms that we use 
to define an ensemble. So, when calculating the mean prediction of the ensemble we use Eq. \eqref{eq:lgpr_c_matrix}
rather than Eq. \eqref{eq:ensemble_mean}. 

The real objective of the ensemble is to
calculate an uncertainty given as the standard deviation of the
predictions, i.e.:
\begin{equation}
	\sigma(X) = \sqrt{\frac{1}{K}\sum_k^K \left(E_k(X) - \bar{E}(X)\right)^2},
	\label{eq:ensemble_unc}
\end{equation}
where $E_k(X) = \sum_i \epsilon_k(\mathbf{x}_i)$, and $\bar{E}_k(X) = \sum_i
\epsilon(\mathbf{x}_i)$, with $\mathbf{x}_i$ being the local descriptors of a
structure with descriptor $X$.

\begin{figure}
	\centering
	\includegraphics[]{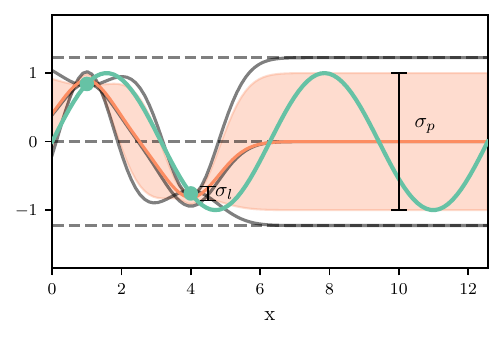}
	\caption{Illustration of an ensemble GPR model consisting of three models,
	the priors $p_k$ of the three models are depicted with \textit{dashed} gray lines, 
	each individual model is a full gray-line. The true function is green, and 
	the mean prediction of the ensemble is orange with the shaded area being the 
	model's uncertainty.}
	\label{fig:ensemble_illustration}
\end{figure}

We illustrate the model in Figure \ref{fig:ensemble_illustration}, from which 
the two new hyperparameters $\sigma_l$ and $\sigma_p$ can be interpreted as the uncertainty at
training points and at points sufficiently far from training data that each 
model just predicts its prior. 

\begin{figure*}
	\centering
	\includegraphics[]{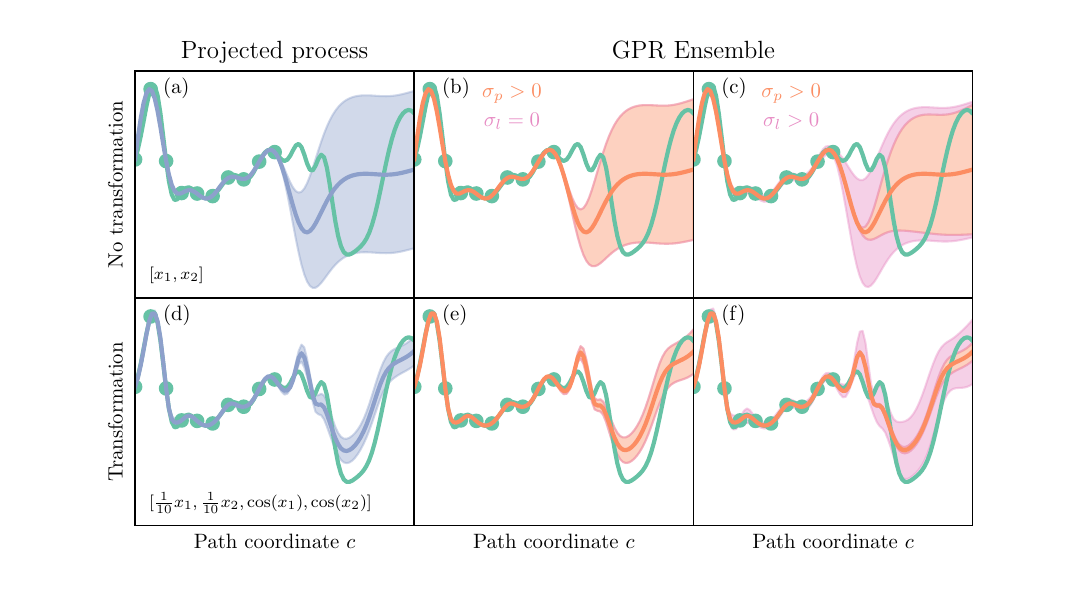}
	\caption[]{GP model fits of the function $f(x_1, x_2) = \cos(x_1) + \frac{x_2 \cos(x_2)}{2}$, 
	along a path. In the top row (a-c) the models use $(x_1, x_2)$ as features, 
	whereas in the bottom row (d-f) $(\frac{1}{10}x_1, \frac{1}{10}x_2, \cos(x_1), \cos(x_2))$
	is used as features exploiting the periodicity of the function to transform the feature space.
	In each row the model fit is the same for every model, the difference lies 
	in the predicted uncertainty. Without a transformation of the feature all three 
	models are very similar and the use of label noise $\sigma_l$ may seem unnecessary.
	However, once the transformation is introduced the longer length-scales 
	that yield superior fits in the extrapolation region the label-noise 
	uncertainty is qualitatively correct.}
	\label{fig:three_models}
\end{figure*}

In figure \ref{fig:three_models} we show three different models, the projected-process sparse GPR, an ensemble GPR 
utilizing only the prior $\sigma_p$ noise to introduce an uncertainty, and a 
ensemble GPR utilizing both the prior $\sigma_p$ and label $\sigma_l$ noise. 
When fitting the potential energy surface descriptors that obey 
translational, rotational and permutational invariance are always used. 
These descriptors encode invariant properties such as bond lengths and angles, 
which correlate distinct atomic configuration. This correlation means 
that a model can and should have low predicted uncertainty about 
configurations that may naively appear to be far from any training example. 
We illustrate the effect of this on the three models in Figure \ref{fig:three_models}(d-f), 
by choosing features that introduce a similar transformation of feature space.
This means that for most kernel length-scales no point is particularly far 
from a training example and as such only a small fraction of $\sigma_p$ can 
ever be realized - whereas even a small $\sigma_l$ can lead to a significant 
uncertainty. 

The form of Eq. \eqref{eq:gpr_ensemble} is advantageous as the training, 
where the majority of the computational expense is computing the matrix $C$, only 
needs to be done once, rather than for each constituent model. Similarly, 
for making predictions the kernel vector $\mathbf{k}_m(\mathbf{x})$ also only 
needs to be calculated once, meaning that predictions for every
constituent model can be made at barely any extra computational cost. 
 
\section{Calibration}
In this section, we compare the quality of the uncertainty prediction
made with the standard closed-form posterior uncertainty expression of
a projected process (see section \ref{sec:method_sgpr} for an introduction) with our proposed ensemble uncertainty method. 
The methodology of calibration curves and errors are presented in section \ref{sec:calibration}

\begin{figure}
	\includegraphics{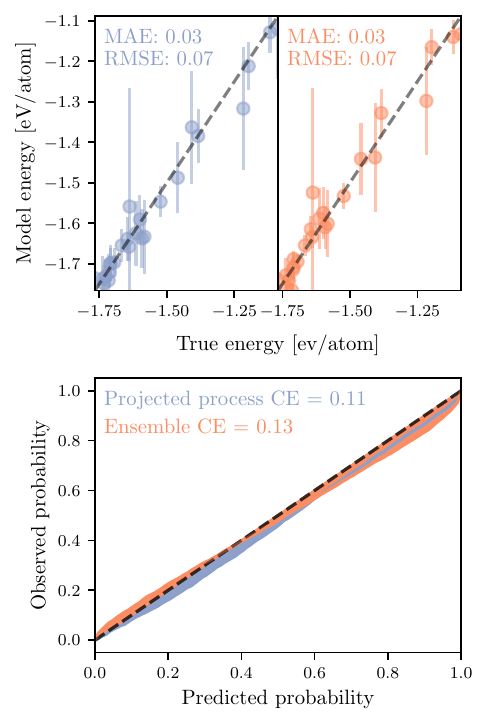}
	\caption{Top left and top right: parity plots with error-bars for the predicted uncertainty for a few randomly selected 
	structures with the regular sparse GPR model and the ensemble model, respectively. MAE and RMSE 
	for the full dataset are reported in eV/atom.	Bottom: 
	Calibration curves for ensemble and regular model in blue and orange, respectively. The values for 
	the calibration error (CE) are given. For the ensemble model,
	Eq.\ \eqref{eq:ensemble_unc} is used for the variance, while for
        the regular model, Eq.\ \eqref{eq:projected_process_posterior_variance}
        is used.}
	\label{fig:au_data_metrics}
\end{figure}

A dataset of Au$_{x}$ with $x=[10, 12, 14, 16, 18, 20]$ consisting of 5838 structures 
has been gathered by selecting structures with unique graphs \cite{christiansenAtomisticGlobalOptimization2022, slavenskyAcceleratingStructureSearch2024} from a set of 
global structure searches for each cluster size. This dataset is split 
into a training set, a validation set and a test set of 225, 25 and 5588
structures respectively. In Fig. \ref{fig:au_data_metrics} we show the 
parity plots for a subset of the test data and calibration plots for both models 
with optimized hyperparameters. Both models are capable of achieving good fits 
with the relatively small amount of training data and both produce well calibrated 
uncertainties. 

For this comparison, we train on the training set and use the validation set 
to calibrate the uncertainties. This calibration amounts to finding 
hyperparameters. From the outset we choose the kernel form (see kernel
definition in Eq. \eqref{eq:kernel}) and its length-scale as $l = 20$.
Having fixed the length-scale, the regular sparse GPR has two hyperparameters 
that can influence the predicted uncertainties, namely the kernel amplitude $\theta_0$
and the noise $\sigma_n$. However, if these are chosen independently the 
model predictions are influenced, to avoid that, we fix the ratio $\frac{A}{\sigma_n} = 100$.
The calibration error can then be minimized wrt. these two parameters while 
keeping the ratio fixed. 

For the ensemble model, we can freely minimize the calibration error wrt. 
to the two additional noise parameters of the ensemble $\sigma_l$ and $\sigma_p$, 
while fixing $\theta_0 = 1$ and $\sigma_n = 0.01$ keeping the same ratio as for 
the regular sparse GPR. 

\section{Global structure search}
To probe the utility of this model we employ it in an active learning global 
structure search algorithm. In each iteration of the algorithm several
structural candidates are stochastically generated and subsequently locally 
optimized in the lower-confidence-bound expression: 
\begin{equation}
	E_\mathrm{LCB}(X) = E(X) - \kappa\sigma(X),
\end{equation}
where $E(X)$ is model total energy prediction, $\sigma(X)$ is the predicted 
standard deviation, and $\kappa$ is a hyperparameter balancing the importance of the 
uncertainty. Among these structural candidates the one with the lowest value of 
$E_{LCB}(X)$ is selected for evaluation with DFT and that structure is added to the 
training set of the model before moving on to the next iteration. The uncertainty 
therefore plays a large role in efficiently exploring the search space. 

\begin{figure}
	\includegraphics[]{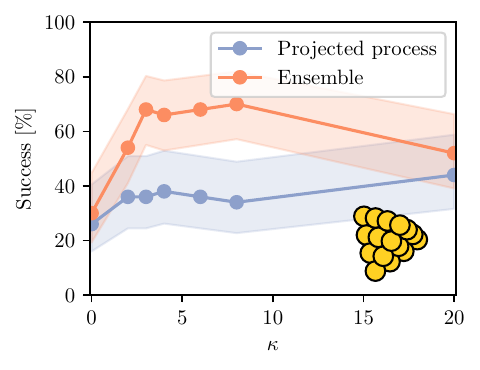}
	\caption{Success as a function of $\kappa$ for Au$_{20}$ for projected process 
	and ensemble models.}
	\label{fig:gold_searches}
\end{figure}

For the largest of the gold clusters, 20 atoms, studied in the previous section we run
many independent searches. For each search we record how many iterations are required in order 
to find the global minimum structure, a perfect tetrahedron, as a function 
of the exploration parameter $\kappa$. If the uncertainty measure is suited for 
this task, there must exist a $\kappa \neq 0$ that increases the number of 
successful searches compared to searches with $\kappa = 0$. In Figure \ref{fig:gold_searches} we present
the success rate, that is the percentage of searches that find the GM structure, as a function 
of $\kappa$ for both the ensemble GPR and the regular GPR using the projected process
expression for the uncertainty. For the ensemble there is a clear improvement in the
search performance as $\kappa$ is increased -- thus the uncertainty helps the 
algorithm explore the configuration space more efficiently.

\section{Timings}
The allure of machine learning models comes in large part from them being 
orders of magnitude faster than traditional quantum mechanical methods, such as DFT. 
Even so, for a task such as the global structure search described in the previous section, 
performing local optimizations in the model constitutes a significant fraction of 
the total time. 

When making a prediction with either type of model the matrix vector of coefficients 
$\alpha = C \mathbf{E}$ can be precomputed for the current training set. 
For the ensemble model this means that the $\alpha$-vector for each model in the ensemble 
can be stored and the time spent calculating the uncertainty from Eq. \eqref{eq:ensemble_unc} 
mainly involves calculating the kernel vectors $\mathbf{k}_m(x)$. 

\begin{figure}
	\centering
	\includegraphics[]{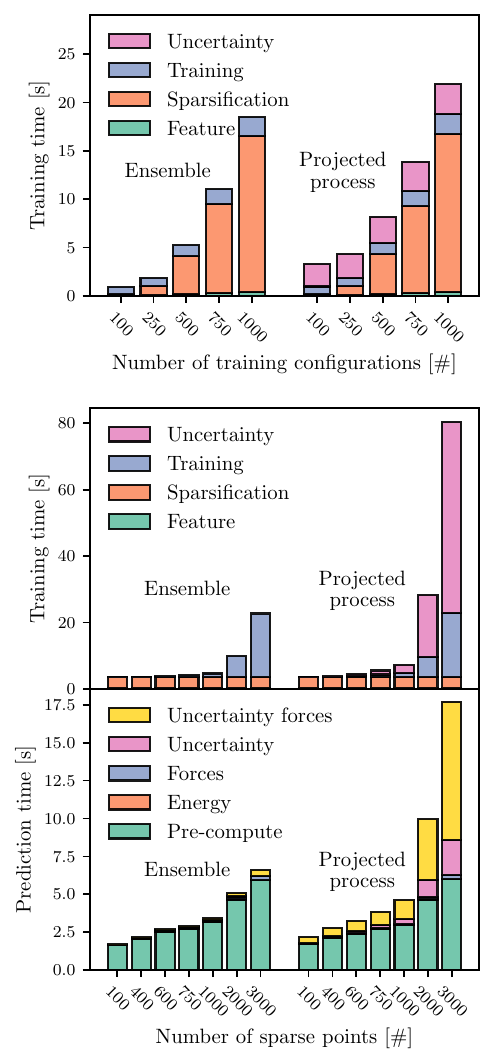}
	\caption{The top and middle panels show timings for different components of training each model and the bottom 
	panel shows timings for making predictions with each model. Timings were done on Intel Xeon Gold 6230 CPUs and averaged 
	over 10 separate runs.} 
	\label{fig:timings}
\end{figure}

Figure \ref{fig:timings} shows timings for training and prediction with both discussed types 
of models on configurations of Au$_{20}$ clusters. Here training covers computing $C\mathbf{E}$ in Eq. \eqref{eq:lgpr_c_matrix} and 
the required matrices for calculating uncertainties, 
such that predictions can be made as fast as possible. As a function of the number of training configurations with a fixed 
number of inducing points the sparsification procedure is the dominating part of the timings, with only a relatively small additional 
time for the projected-process model as it involves inverting additional matrices. With a fixed set of training data 
but a varying number of inducing points a more pronounced difference between the two methods can be observed, 
which again can be attributed to the projected-process uncertainty expression needing the construction and inversion 
of additional matrices that grow with the number of inducing points. Finally for predictions on 50 configurations, for the ensemble the timings 
are dominated by the pre-computing features and kernel elements between the query configuration and the set of inducing points.
Whereas, for larger sets of inducing points the calculation of uncertainties and especially derivatives of the 
uncertainties in the projected process starts becoming a significant proportion of the time. 

\section{Disucssion}
\label{sec:disucssion}

We have investigated the efficacy of two expressions for predicting uncertainties of 
GPR models in the realm of atomistic simulations. It has been shown that both 
models can produce well-calibrated uncertainties. However, in a common material 
science simulation task, namely global structure search for atomic configurations we 
have found that the proposed ensemble GPR uncertainty expression is advantageous. 
Further, we document that the ensemble method has superior behavior in terms of 
computational cost -- both when it comes to training and prediction. 

\section{Methods}
\label{sec:methods}

\subsection{Sparse Gaussian Process Regression}
\label{sec:method_sgpr}

We have previously reported on a sparse local GPR \cite{ronneAtomisticStructureSearch2022}, where the covariance between 
local enviroment descriptors is given by
\begin{equation}
	k(\mathbf{x}_i, \mathbf{x}_j) = \theta_0 \exp\left(-\frac{|\mathbf{x}_i-\mathbf{x}_j|^2}{2l^2} \right).
	\label{eq:kernel}
\end{equation}
Here, $\theta_0$ and $l$ are amplitude and length-scale hyperparameters of this radial basis 
function kernel. With a training dataset $X_n$ and a set of inducing points $X_m$, 
we can define covariance matrices $K_{mm}$ and $K_{nm}$ as the covariance between all 
descriptors in $X_m$ and $X_m$ and $X_n$ and $X_m$. From that, we can define
\begin{equation}
	C = \left[K_{mm} + (LK_{nm})^T \Sigma_{nn}^{-1}(LK_{nm})\right]^{-1}(LK_{nm})^T\Sigma^{-1},
\end{equation}
where $\Sigma_{nn}$ is a diagonal matrix with the noise of each environment and
a local energy may be predicted using 
\begin{equation}
\epsilon(\mathbf{x}) = \mathbf{k}_m(\mathbf{x})C \mathbf{E}.
\label{eq:sparse_lgpr_energy}
\end{equation}
Here $\mathbf{E}$ are the observed total energies and $L$ is a local energy correspondence matrix. 
Compared to a non-sparse GPR model, which is recovered in the limit where 
$X_m = X_n$, the sparse GPR comes at reduced computational cost as the matrix 
that is inverted is only of size $m \times m$ and the vector of kernel 
elements between a query-point and $X_m$ does not grow with the training set 
size. From an intuitive point of view this sparsification also necessitates a change to 
the predicted variance. In the projected process form the covariance
is given by \cite{rasmussenGaussianProcessesMachine2006}:
\begin{equation}
	\begin{aligned}
		&\varsigma(\mathbf{x}_i, \mathbf{x}_j)^2 = k(\mathbf{x}_i, \mathbf{x}_j) -  \mathbf{k}^T_m(\mathbf{x}_i)K_{mm}^{-1}\mathbf{k}_m(\mathbf{x}_j) \\ 
		&+\mathbf{k}_m(\mathbf{x}_i)^T (K_{mm} + K_{mn}\Sigma^{-1}_{nn}K_{nm})^{-1}\mathbf{k}_m(\mathbf{x}_j).
	\end{aligned}
	\label{eq:pp_variance}
\end{equation}
Here  $\varsigma(\mathbf{x_i}, \mathbf{x_j})^2$ is the predicted covariance of two local energies.
As $\Sigma_{nn}$ involves every local environment in the training set it can become quite large, so unless a 
sparse matrix implementation is used that can lead to memory issues. Further, the matrix product 
involving $\Sigma_{nn}$ is also much faster using a sparse matrix implementation.
For the acquisition function considered in this work we are interested in the standard 
deviation of the total energy
\begin{equation}
	\sigma(X) = \sqrt{\sum_{\mathbf{x}_i \in X} \sum_{\mathbf{x}_j \in X} \varsigma(\mathbf{x}_i, \mathbf{x}_j)^2}
\label{eq:projected_process_posterior_variance}
      \end{equation}
that takes into account the covariances of the local energies. 

\subsection{CUR Sparsification}
In order to choose the local environments in the set of inducing points $X_m$ we employ 
the CUR algorithm \cite{mahoneyCURMatrixDecompositions2009}. Here the full feature matrix $X_n$, that has dimensions $(n, f)$ where $n$ is
the number of local environments and $f$ is the number of features, is decomposed by singular value 
decomposition into three matrices $X_n = U\Sigma V$, where $\Sigma$ is
a diagonal matrix with entries in descending order. A probability of selection is then calculated for 
each local enviroment, indexed by $i$, as
\begin{equation}
	P_i \propto \sum_{j<k} {U_{ij}^2}
\end{equation}
where $k = \min(n, f)$ which for all but the smallest datasets is equal to $f$. 
A predetermined number $m$ of environments is finally picked, based on the calculated 
probabilities, in order to establish the set of inducing points $X_m$. 

\subsection{Calibration}
\label{sec:calibration}
In the context of uncertainty quantification calibration refers to ensuring that the 
true labels fall within a certain confidence interval given by the predicted standard 
deviation. To this end we use calibration plots, as introduced by \cite{Kuleshov2018}, to asses the quality of the uncertainty estimate 
of a given model. When recalibrating models we minimize their calibration error, also defined by \cite{Kuleshov2018}.
These metrics have been employed by other authors in the area of MLIPs \cite{tran_methods_2020,hu_robust_2022}.

\section{Code availability}
The code used the findings presented in this paper is publicly available as part of AGOX as of version 2.7.0 at https://gitlab.com/agox/agox under a GNU GPLv3 license. 

\section{Acknowledgements}
We acknowledge support from VILLUM FONDEN through Investigator grant, project no. 16562, and by the Danish National Research Foundation through the Center of Excellence “InterCat” (Grant agreement no: DNRF150).

\bibliographystyle{apsrev4-1}
\bibliography{references}

\end{document}